\documentclass[12pt]{article}
\usepackage[inner=1.25in,outer=1in,bottom=1.1in, top=1.25in]{geometry} 
\usepackage{fontspec}
\usepackage{newtxtext} 
\usepackage{setspace}
\usepackage{ragged2e} 
\RaggedRight 
\usepackage[none]{hyphenat} 
\usepackage{titlesec}
\usepackage{enumitem}
\usepackage{hyperref}
\usepackage{placeins}

\usepackage{graphicx} 
\usepackage{amsmath}
\usepackage{bm}
\usepackage{dsfont}
\usepackage{threeparttable}
\usepackage{tabularx} 
\usepackage{booktabs}
\usepackage{caption}
\usepackage{float}
\usepackage[T1]{fontenc}
\usepackage[
backend=biber,
style=apa,
style=nature, 
maxcitenames=99, 
uniquelist=false 
]{biblatex}

\usepackage{authblk}

\title{Bayesian Geostatistical Modeling for Cluster Randomized Trials}
\author[1]{Jooyeon Lee, M.S.}
\author[1]{Evan Kwiatkowski, Ph.D.$^{*}$}
\affil[1]{Department of Biostatistics \& Data Science, UTHealth Houston School of Public Health, Houston, TX, USA\\
$^{*}$Corresponding author: Evan.K.Kwiatkowski@uth.tmc.edu}
\date{}

\begin{document}
\maketitle

\begin{abstract}

\end{abstract}
Cluster randomized trials (CRTs) offer a practical alternative for addressing logistical challenges and ensuring feasibility in community health, education, and prevention studies, even though individually-randomized controlled trials are considered the gold standard in evaluating therapeutic interventions. Despite their utility, CRTs are often criticized for limited precision and complex modeling requirements. Advances in robust Bayesian methods and the incorporation of spatial correlation into CRT design and analysis remain relatively underdeveloped. This paper introduces a Bayesian geostatistical framework that models individuals nested within geographic clusters while explicitly accounting for spatial dependence. We demonstrate that conventional non-spatial models are susceptible to underestimating uncertainty and lead to misleading inferences, whereas our spatial approach improves estimation stability, controls type I error, and enhances statistical power. Additionally, we explore design implications that are suggested through the exploration of spatial predictive uncertainty. Our results of simulation and real-world data application demonstrate the value and need for wider adoption of spatial methods in CRTs.


\newpage 

\section{Introduction}
Individually-randomized controlled trials are represented as the gold standard in evaluating healthcare interventions \cite{schulzCONSORT2010Statement2010}. However, in many clinical and public health studies, randomizing individuals is often impractical due to the logistical complexity of delivering interventions and the risk of treatment contamination among the individuals. Additionally, an individually randomized controlled trial may be unacceptable to the local population, which could threaten a trial’s successful completion and its social and scientific value that is essential to its justification \cite{whoEthicalIssuesRelated2014, kennedyImplementationEbolaVirus2016, kahnChoicesVaccineTrial2018}. Furthermore, for research intended to inform policy reform, cluster randomized designs may provide findings that generalize better to real-world policy contexts \cite{raudenbushStatisticalAnalysisOptimal}. 
 
 In such cases, cluster randomized trials (CRTs) can provide a robust alternative in which the interventions are randomized by groups of individuals, either by geography or institution to evaluate the effectiveness of interventions. CRTs have become a fundamental design in evaluating interventions in community health, education, and health service delivery and have been a recurrent theme in clinical trial designs with methodological research devoted to matched paired designs, intracluster correlation coefficient (ICC), and sample size calculations to address issues of clustering and power \cite{campbellDevelopmentsClusterRandomized2007}. Examples of CRTs in health care settings include primary care clinics of hospitals \cite{boultEffectGuidedCare2011, moherClusterRandomisedControlled2001}, schools or community-based programs, clinical decision support systems for mental health training \cite{benrimohAifredHealthDeep2018, tanguay-selaEvaluatingPerceivedUtility2022}, and regional vaccination campaigns \cite{borgeyEffectivenessInterventionCampaign2019, banerjeeImprovingImmunisationCoverage2010, varmaResearchProtocolTwo2019}.

Parallel CRTs are the most common clinical trial design that randomizes clusters to one of two or more arms \cite{lewisStatisticalPrinciplesClinical1999}. The randomizing unit in this design is a cluster, so all individuals in each cluster receive only one kind of treatment during the study, usually starting simultaneously. There are several methods for analyzing parallel CRTs. In a case where cluster sizes are equal and in a two-arm trial, a simple way is to use a two-sample t-test to compare cluster-level mean responses between the groups. When there are more than two treatment arms, a one-way analysis of variance may be applied. In a special case, where a matched-pairs design is used, a paired t-test can be conducted. When cluster sizes are unequal, individual-level analysis such as random effects models \cite{lairdRandomeffectsModelsLongitudinal1982} or generalized estimating equations (GEE) \cite{liangLongitudinalDataAnalysis1986} may be used. 

A defining feature of CRTs is the ICC, which quantifies the degree of similarity in outcomes among individuals within the same cluster. A high ICC implies that individual responses are highly correlated, which must be accounted for at both the design and analysis stages. Ignoring ICC can lead to underpowered studies and inflated type I error rates. Standard analytic strategies include mixed-effects models, which explicitly model within‑cluster correlation through random effects, and GEE, which accounts for within‑cluster correlation via a specified working correlation structure \cite{hayesClusterRandomisedTrials2009}. The Consolidated Standards of Reporting Trials (CONSORT) \cite{campbellConsort2010Statement2012} statement provided reporting guidelines for CRTs to include ICC for each primary outcome and account for the design effect induced by ICC in sample size calculations.

Bayesian methods in CRTs allow the incorporation of external evidence by using substantive prior opinions on key parameters, such as treatment effects, cluster-level variances, and intraclass correlations. They also provide a flexible framework for modeling hierarchical structures and capturing dependence among responses or random effects. In addition, uncertainty about model assumptions can be directly incorporated into power calculations. \textcite{spiegelhalterBayesianMethodsCluster2001a} showcased Bayesian approaches for CRTs with continuous responses and highlighted that considerable development of robust strategies for Bayesian modeling in CRTs is still needed. A methodological systematic review of Bayesian methods in CRTs noted the lack of application of Bayesian methods relative to frequentist approaches and underscored the need for further Bayesian methodological developments in both the design and analysis of CRTs \cite{jonesBayesianStatisticsDesign2021}. 

Geographic clusters are the most commonly chosen randomization unit for trials in which subgroups or entire populations are targeted for intervention testing \cite{hayesClusterRandomisedTrials2009}. Examples 
include insecticide bed-net distribution \cite{gimnigEFFECTPERMETHRINTREATEDBED2003, binkaImpactSpatialDistribution1998, hawleyCOMMUNITYWIDEEFFECTSPERMETHRINTREATED2003, lenhartInsecticidetreatedBednetsControl2008}, water sanitation programs \cite{freemanAssessingImpactSchoolbased2012}, vector-control interventions \cite{alexanderSpatialVariationAnophelestransmitted2003, kroegerEffectiveControlDengue2006}, vaccination/health campaigns \cite{aliGeographicAnalysisVaccine2007, miguelWORMSIDENTIFYINGIMPACTS2004}, and vaccine effectiveness study \cite{chaoContributionNeighboursIndividuals2015}, where outcomes in neighboring areas are inherently correlated. Especially in vaccine trials, spatial dependency often arises from indirect protection mechanisms such as herd immunity \cite{johnHerdImmunityHerd2000} or human mobility patterns \cite{zhuSpacetimeRelationshipsCOVID192023}. A systematic review of spatial CRTs \cite{jarvisSpatialAnalysisCluster2017} discussed that the incorporation of spatial effects in CRTs is rare and called for further development and evaluation of spatial methodologies across a range of CRT designs. 

There have been a few attempts to quantify the importance of spatial dependency in trials through spatial correlation \cite{barriosClusteringSpatialCorrelations2012} and spatial lag models \cite{baylisHowImportantSpatial2015}. \textcite{guoBayesianAdaptivePhase2019} extended the trial design to account for spatial variation through a Bayesian adaptive design using an area-level conditional autoregressive prior. Spillover effects in a CRT of insecticide-treated nets (ITNs) were analyzed through the use of conditional autoregressive models and Gaussian process \cite{JarvisSpatialEffects2019, jarvisSpatialAnalysisCluster2018}. However, these studies primarily focused on secondary spatial analyses of existing CRT data and estimation of spillover effects, rather than systematically examining the performance metrics of treatment effect estimation under varying ICC and spatial covariance regimes.

Spatially structured CRTs are particularly important when outcomes or treatment effects vary over space due to underlying geographic heterogeneity in environmental exposures, health system infrastructure, socioeconomic context, or other unmeasured confounders. While CRTs have received considerable attention in the literature, most analytic methods have not incorporated spatial considerations in design or analyses, even though the need for spatial modeling arises naturally in many cluster-based interventions. Systematically evaluating the performance under realistic ICCs and various spatial structures is important because they jointly influence variance and treatment effect inference, which directly affects trial design, sample size determination, and statistical validity. However, systematic evaluations of CRTs that leverage point-referenced participant addresses remains limited, and to our knowledge, the potential role of cluster-level spatial predictive uncertainty has not been explored.

In this paper, we propose a Bayesian geostatistical method in CRTs that accommodates spatial variation of participants' geocoded locations and individual-level covariates and cluster-level features. The objective of the proposed method is to assess whether the intervention treatment is superior to the control in various realistic ICC and spatial settings. We account for both the cluster-specific variations and spatial variation, where we introduce a Gaussian process with a geostatistical structure for spatial dependency. Our proposed method adopts an individual-level approach to analyze CRTs that accounts for spatial structure. We compared this model with individual and cluster-level models that ignore spatial effects, evaluating performance using metrics such as power, false positive rate, percent relative error, mean squared error, bias, and coverage rate for the treatment effect.    

The primary contribution of this paper lies in the rigorous evaluation of a geostatistical mixed-effects framework within the CRT inference setting. Specifically, our contribution is twofold: first, we adopt a geostatistical mixed-effects model tailored to CRT settings and systematically evaluate the impact of incorporating spatial correlation on treatment effect estimation through simulation studies, demonstrating improvements in power, bias, and precision compared with non-spatial CRT analysis methods; second, we provide a practical Bayesian implementation using the Integrated Nested Laplace Approximation (INLA) \cite{rueApproximateBayesianInference2009a}. Additionally, we examined cluster-level spatial predictive uncertainty to illustrate the proposed model's potential for informing future treatment allocation strategies.

In Section \ref{sec:methods}, we present a Bayesian geostatistical method for treatment efficacy trials and its potential use in CRTs with continuous responses. Section \ref{sec: design implications of spatial uncertainty} illustrates potential design implications of treatment allocation strategies from summarized spatial predictive uncertainty. Section \ref{sec:simulation} outlines simulation settings with data generation, performance metrics, nonspatial comparison methods to compare the performance with our proposed method, and simulation results. Section \ref{sec:data applcation} shows the application of methods in clustered data, and lastly, Section \ref{sec:discussion} describes the advantages and disadvantages of the methods covered in this paper. Additional details are provided in the Supplementary Materials. 

\section{Methods}\label{sec:methods}
\subsection{CRT design and Gaussian process}\label{sec:CRT AND GAUSSIAN PROCESS}
Consider a two-arm CRT, where $z_i \in \{0,1 \}$ for cluster $i$ denotes the treatment arm, control and intervention, respectively. Let $\mathcal{I}$ denote the set of cluster regions and let $|\mathcal{I}|=I$ be the number of clusters. Without loss of generality, each cluster $i \in \mathcal{I}$ indexed by $i=1,...,I$ consists of $J$ individuals indexed by $j=1,...,J$. Thus, the total number of participants in the trial is $N = IJ$. Since we are conducting a cluster randomized trial (CRT), in which randomization occurs at the cluster level by definition, we assume that all individuals within a given cluster are assigned to the same arm, either intervention or control. This design choice is consistent with best practices in CRTs, which helps prevent contamination between arms and simplifies trial implementation \cite{hayesClusterRandomisedTrials2009}. 

We specifically consider the geographical type of cluster, which could be based on administrative units, any other type of geographic zones, institutional clusters such as schools, health units, or workplaces. Let $Y_{ij}$ denote the continuous outcome for the individual $j$ in cluster $i$, located at spatial coordinate $\bm{s}_{ij} \in \mathcal{B}_{i} \subset \Omega$, where a global spatial domain of the trial is denoted by $\Omega \subset \mathds{R}^2$ and $\mathcal{B}_i$ is the spatial region associated with cluster $i \in \mathcal{I}$.

For any finite set of spatial locations $\{ \bm{s}_{ij}: i=1,...,I; j=1,...,J \}$, if the outcome vector $\bm{Y} = (Y_{11},...,Y_{1J},..., Y_{I1},..., Y_{IJ})'$ follows a multivariate normal distribution with mean vector $\bm{\mu} = (\mu_{11},...,\mu_{1J},..., \mu_{I1},..., \mu_{IJ})'$ and covariance matrix $\bm{\Sigma}$, then the collection $\{ Y(\bm{s}_{ij}): \bm{s}_{ij} \in \Omega \} $ is a spatial stochastic process and is said to be a Gaussian process (GP). The generic element of $\bm{\Sigma}$ is defined by a covariance function $C(.,.)$ such that 
$\Sigma_{(ij),(i'j')} = \text{cov}(y(\bm{s}_{ij}), y(\bm{s}_{i'j'})) = C(\bm{s}_{ij}, \bm{s}_{i'j'})$ \cite{banerjeeHierarchicalModelingAnalysis2003a, blangiardoSpatialSpatiotemporalBayesian2015}. We assume the covariance function to be weakly stationary, in which the covariance between outcomes depends only on the distance between locations. Although our setup is for continuous outcomes, it is also applicable to other data types, such as binary outcomes, as demonstrated in the real-data example in Section \ref{sec:data applcation}.

A hierarchical mixed-effects model incorporating within-cluster variance $\sigma_W^2$ and between-cluster variance $\sigma_{B}^2$ is considered to account for individual-level and cluster-level variability in CRTs and is well described in Hayes and Moulton \cite{hayesClusterRandomisedTrials2009}. We extended unobserved variability to the spatial domain using a geostatistical model where the residual is partitioned into two components: a spatially structured Gaussian process and an independent random noise term. The latent spatial component captures underlying spatial patterns that are not explained by cluster-specific effects or random noise \cite{wikleSpatiotemporalStatistics2019a}.

\subsection{Proposed method: CRT-SMM} \label{sec:PROPOSED DESIGN}
We introduce a CRT spatial mixed-effects model (CRT-SMM) that offers a novel application of spatial modeling in the context of cluster randomized trials, which accounts for both between-cluster variance and geostatistical correlations. We assumed that geostatistical correlation is shared across the entire region of interest, thus the outcome $Y_{ij}$, observed at spatial location $\bm{s}_{ij}$, is modeled using a mixed-effects framework continuously defined over a spatial domain $D \subset \mathds{R}^2$, and is formulated as 
\begin{align} \label{eq:y_s}
    Y(\bm{s}_{ij}) = \mu_{ij} + u_{i} + w(\bm{s}_{ij}) + \epsilon_{ij},
\end{align} 
where $\mu_{ij}$ is an individual-level mean structure that accounts for cluster-level treatment indicator $z_{i} \in \{0,1\}$; and a vector of $K$ individual-specific covariates $\bm{x}_{ij} = (x_{ij0}, x_{ij1},...,x_{ij(K-1)})'$ such as demographic, socioeconomic, or biomarker variables. The covariate term $x_{ij0}$ is set as 1 to reflect an intercept. The individual-level mean structure $\mu_{ij}$ is defined as follows:   
 \begin{align}\label{eq:mu}
     \mu_{ij} =  \beta z_{i} + \bm{x}_{ij}'\bm{\gamma} + z_{i} \cdot \bm{x}_{ij}' \bm{\delta},
\end{align}
where $\beta$ is fixed treatment-specific effects, $\bm{\gamma} \in \mathds{R}^K$ is fixed covariate effects, and $\bm{\delta} \in \mathds{R}^K$ is a fixed effects interaction term between treatment and individual-specific covariates that accounts for treatment effect heterogeneity. Note that the cluster-specific treatment indicator $z_i$ is used instead of the individual-level $z_{ij}$ since randomization was conducted at the cluster level although this assumption can be relaxed as demonstrated in the real data example. 

 In this model, $u_{i}$ denotes a cluster-specific random effect that is assumed to independently and identically (i.i.d.) follow a normal distribution with mean zero and has between-cluster variance $\sigma_{B}^2$. Individual-level random noise $\epsilon_{ij}$ is assumed to i.i.d. follow a normal distribution with mean zero and within-cluster variance $\sigma_W^2$. Spatial random effects denoted by $w(\bm{s}_{ij})$ are defined over individual-level spatial locations $\bm{s}_{ij}$. The three components $u_i$, $\epsilon_{ij}$, $w(\bm{s}_{ij})$ are assumed to be mutually independent of each other. 
 
 The collection of spatial effects $\{w(\bm{s}_{ij}): \bm{s}_{ij} \in D \}$ is assumed to follow a zero-mean Gaussian Process (GP) with the covariance function:
 $$
 \text{cov}(w(\bm{s}_{ij}), w(\bm{s}_{i'j'})) = \tau^2 \cdot \rho(\bm{s}_{ij}, \bm{s}_{i'j'}),
 $$
 where $\tau^2$ is the marginal spatial variance and $\rho(.,.)$ is an isotropic decay function that depends only on the distance between locations. See Section A1 of Supplementary Materials for the covariance structure of the outcome vector marginalized over spatial effects $w(\bm{s}_{ij})$. 

The ICC defined in a model with spatial dependency is
 \begin{align*}
     \text{ICC} = \frac{\sigma_B^2 }{\sigma_B^2 + \tau^2 + \sigma_W^2}.
 \end{align*}

 The resulting marginal likelihood for all individuals in the study is defined through $N$-dimensional outcome vector $\bm{Y}$, where $N=IJ$. The outcome vector follows a multivariate normal distribution as follows: 
 \begin{align*}
     \bm{Y}|\beta, \bm{\gamma}, \bm{\delta}, \tau^2, \phi, \sigma_{B}^2, \sigma_{W}^2 \sim MVN(\bm{\mu}, \ \bm{\Sigma}),
 \end{align*}
 where the mean vector and covariance matrix is specified as 
 \begin{equation}\label{eq:mu_sigma}
     \begin{aligned}
     \bm{\mu} &= \bm{z}\beta + \bm{X}\bm{\gamma}+ diag(\bm{z}) \cdot \bm{X}\bm{\delta}, \\
     \bm{\Sigma} &= \tau^2 \bm{H}(\phi) + \sigma_{B}^2\bm{C}^T\bm{C} + \sigma_{W}^2 \bm{I}.
     \end{aligned}
 \end{equation}
The elements in the treatment assignment vector $\bm{z} \in \mathds{R}^{N \times 1}$ contain binary indicators for each individual, where $z_{ij}$ is 1 if individual $j$ in cluster $i$ is assigned to the intervention, and 0 if assigned to control group. A cluster membership matrix $\bm{C} \in \mathds{R}^{I \times N}$ and spatial correlation matrix $\bm{H}(\phi) \in \mathds{R}^{N \times N}$ are presented in detail Section A1 of Supplementary Materials. The spatial range parameter $\phi$, also referred to as a decay parameter, controls the reach of the spatial distance (i.e., larger values indicating greater spatial correlation). In a CRT framework, the elements of $\bm{z}$ are identical for all individuals within the same cluster, which means $z_{ij}=z_i$ for all $j$ in cluster $i$. Thus, while the notation $\bm{z}$ is defined at the individual level, its values reflect cluster-level randomization. $\bm{X} \in \mathds{R}^{N \times K}$ is a matrix of $K$ individual-level covariates; $\bm{\gamma}, \bm{\delta} \in \mathds{R}^{K \times 1}$ are main effects of covariates and treatment-covariate interaction effects, respectively.

CRT-SMM can be summarized by adopting the recast of a geostatistical specification to a hierarchical form from \cite{gelfandSpatialStatisticsGaussian2016}:
\begin{align}\label{eq:hierarchical}
\begin{split}
    Y_{ij}|\beta, \bm{\gamma}, \bm{\delta},  \bm{u}, \bm{w}(\bm{s}),  \sigma_{W}^2 &\sim N ( \mu_{ij}  + u_i + {w}(\bm{s}_{ij}), \ \sigma_{W}^2)\\
    \bm{w}(\bm{s})| \tau^2, \phi &\sim GP(\bm{0}, \tau^2\bm{H}(\phi)),
\end{split}
\end{align} 
where $\bm{w}(\bm{s}_{ij})$ denotes the realization of a latent spatial process $\bm{w}(\bm{s}) \in \mathds{R}^{N \times 1}$ at the location $\bm{s}_{ij}$ of individual $j$ in cluster $i$. The spatial process $\bm{w}(\bm{s}) $ is modeled as a zero-mean Gaussian process with covariance structure governed by range parameter $\phi$ and marginal variance $\tau^2$. 

We assume that a larger value is preferred for continuous outcomes, and are interested in evaluating the intervention by testing the hypothesis
\begin{align}\label{eq:hypothesis}
    H_0: \theta \leq \Delta \ \text{vs.} \ H_1: \theta > \Delta,
\end{align}
where $\theta$ is an overall (i.e., marginal) treatment effect denoted by the difference of the treatment arms and $\Delta$ is a minimum clinically important difference. Specifically, $\theta = \mu_1 - \mu_0$, with the population-average (i.e., marginal) treatment effect estimator defined as  
\begin{align*}
    \hat{\mu}_1 &= E[y_{ij}|z_{i}=1] = \hat{\beta} + E[\bm{x}_{ij}|z_{i}=1]'(\hat{\bm{\gamma}} + \hat{\bm{\delta}})\\
\hat{\mu}_0 &= E[y_{ij}|z_{i}=0] = E[\bm{x}_{ij}|z_{i}=0]'\hat{\bm{\gamma}},
\end{align*}
where each of $\mu_0$ and $\mu_1$ represents the expected treatment response measure in the control and intervention arms, respectively. When the covariates are centered within each arm or cluster, then $\theta$ is simplified to $\theta = \beta$ which captures the average treatment effect, free from the influence of covariates. 

In the Bayesian framework, a closed form of the posterior cannot be derived from our proposed model due to the hierarchical structure and the presence of the spatially correlated Gaussian process (GP) prior. Moreover, the Markov chain Monte Carlo method (MCMC) requires matrix inversion of the GP covariance matrix $\bm{\Sigma}$, leading to computational complexity. We thereby adopted INLA by leveraging the stochastic partial differential equations (SPDE) approach \cite{lindgrenExplicitLinkGaussian2011} and obtained posterior approximates for the treatment effect.  

\section{Design implications of spatial predictive uncertainty} \label{sec: design implications of spatial uncertainty}

\subsection{Spatial predictive uncertainty and estimation efficiency}
In spatial statistics literature, optimizing data collection based on previous data is referred to as adaptive geostatistical design \cite{chipetaAdaptiveGeostatisticalDesign2016}. For example, \cite{diggleBayesianGeostatisticalDesign2006} proposed a model-based geostatistical approach for choosing sampling locations that minimize the integrated prediction variance of a spatial random field, with the goal of efficient spatial prediction. While our scientific objective differs from that of efficient geostatistical prediction, we adopt the same underlying principle: reducing predictive uncertainty in spatial random effects may reduce variability in treatment effect estimates and potentially inform allocation strategies. 

The proposed CRT-SMM framework yields cluster-specific measures of spatial predictive uncertainty through the averaged prediction variance (APV) of the latent spatial field $w(\bm{s})$. For cluster $i$ with geographic domain $\mathcal{B}_i \subset \Omega$, we define
\begin{align*}
    \text{APV}_i = \int_{\mathcal{B}_i} \text{Var}(w(\bm{s}) \mid \bm{D}) d\bm{s},
\end{align*}
where the data $\bm{D} = \{ Y(\bm{s}_{ij}), \bm{x}_{ij}, z_i; \ i=1,...,I, j=1,...,J \}$. The APV design criterion is adopted from \textcite{diggleBayesianGeostatisticalDesign2006}, which quantifies spatial predictive uncertainty through marginal posterior variances. Although the criterion is expressed in terms of marginal variances, these are computed under the full spatial covariance model and therefore implicitly incorporate the underlying spatial dependence structure, as demonstrated in Section A2 of the Supplementary Materials. 

Under heterogeneous spatial density or varying geographic dispersion of participant locations, for example, clusters located in rural versus urban settings, $\text{APV}_i$ may differ substantially across clusters. Such heterogeneity implies that equal allocation across clusters may not be information-wise efficient, as clusters with higher predictive uncertainty may contribute more to variability in treatment effect estimates.

We formally demonstrate how cluster-level APV relates to the precision of the treatment effect estimator. The Fisher information quantifies the information that the sample carries about the unknown parameter. The Fisher information for the treatment effect is derived as $\mathcal{I}(\theta) = \bm{z}' \bm{\Sigma}^{-1}\bm{z}$ in the Section A2 of Supplementary Materials, and can be approximated by 
\begin{align}\label{eq:approx fisher info}
    \sum_{i=1}^I \bm{z}_i' \bm{\Sigma}_i^{-1} \bm{z}_i = \sum_{i=1}^I \frac{J z_i^2}{\sigma_W^2 + \sigma_B^2 + \text{APV}_i}.
\end{align}
The approximation in equation (\ref{eq:approx fisher info}) is valid when between-cluster spatial correlation is weak, such as in settings where clusters are geographically well separated. In this case, the covariance matrix $\bm{\Sigma}$ can be reasonably approximated by a block-diagonal structure.

\subsection{Illustrative example under heterogeneous cluster density}
\begin{figure}[h]
    \centering
    \includegraphics[width=0.9\linewidth, trim=0 150 0 150]{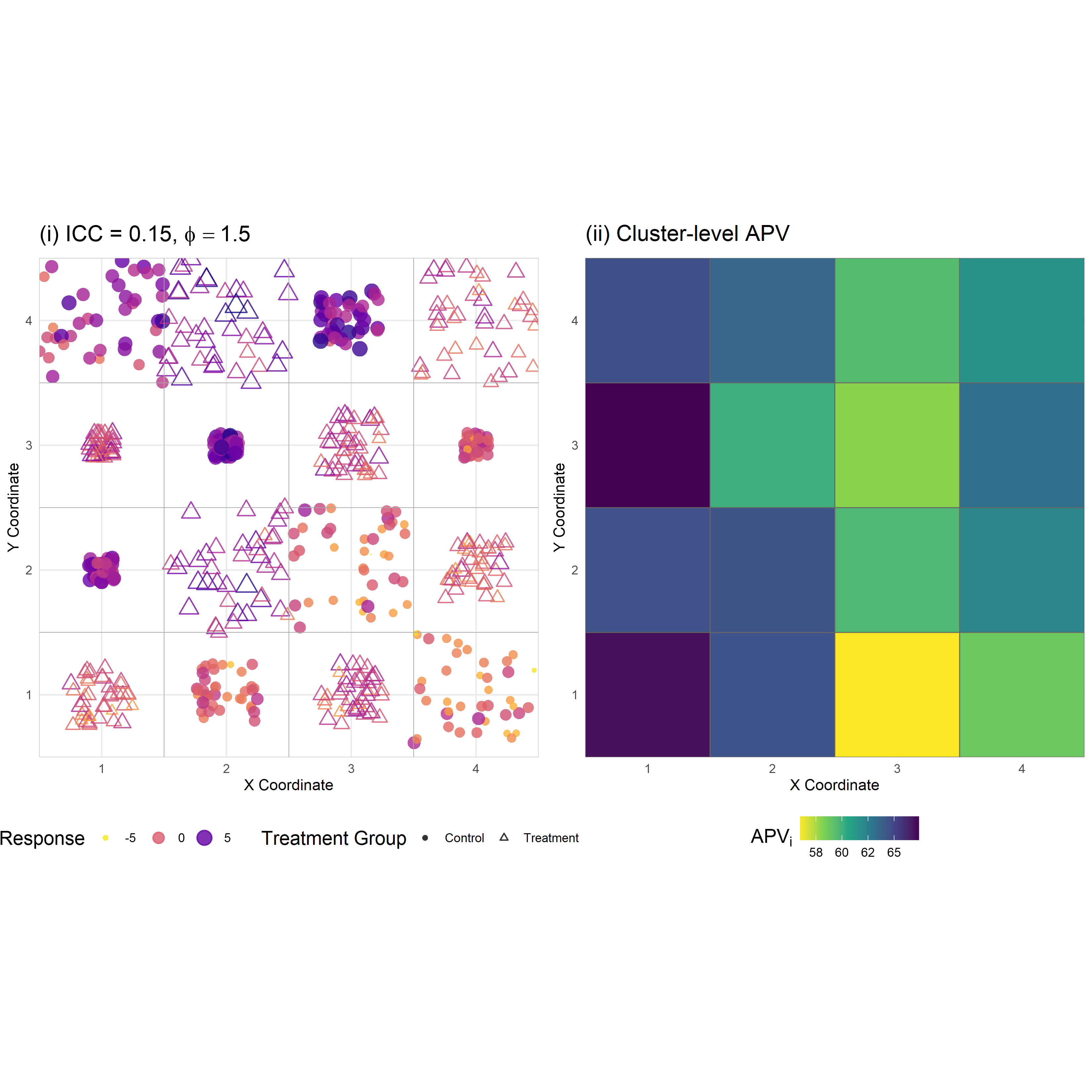} 
    \caption{An illustrative example of cluster-level $\text{APV}_i$ when heterogeneous density of individuals in cluster exists. This is one example when $\text{ICC}=0.15$, $\phi=1.5$, and when true treatment effect is $\theta=0.5$.}
    \label{fig:fig5}
\end{figure}

As an illustrative example, we examined the distribution of $\text{APV}_i$ across clusters under heterogeneous spatial density, specifically in the case of moderate ICC ($\text{ICC}=0.15$), $\phi=0.15$, and when the true treatment effect is $\theta=0.5$ (i.e., Scenario C of Section \ref{sec:data generating mechanism}). Figure  \ref{fig:fig5} shows the distribution of $\text{APV}_i$ across clusters under the checkerboard randomization structure. We further assessed the association between cluster-level predictive uncertainty and the variance of the estimated treatment effect, which is depicted in Figure S1 of the Supplementary Materials.

The clusters with larger $\text{APV}_i$ are associated with higher treatment effect variability, suggesting that recruitment strategies accounting for spatial predictive uncertainty may reduce sampling error and improve precision, especially under heterogeneous cluster densities. This finding motivates future investigation on adaptive allocation strategies that prioritize enrollment in clusters with higher predictive uncertainty, although the formal development and evaluation of such procedures are beyond the scope of the present study.

\section{Simulation studies}\label{sec:simulation}
\subsection{Data generation} \label{sec:data generating mechanism}
We conducted simulation studies to evaluate the performance of CRT-SMM in comparison with nonspatial methods described in Section \ref{sec:comparison methods}, under realistic CRT scenario settings. In our simulation, we focused on a 1:1 geographical randomization scheme to ensure an acceptable overall balance in treatment allocation. For CRTs, it is generally recommended to have highly restrictive randomization balanced on individual-level covariates or cluster-level when there is a limited number of clusters \cite{raabBalanceClusterRandomized2001, moultonCovariatebasedConstrainedRandomization2004}. Geographic balance is especially important to reduce the likelihood that observed differences are driven by spatial heterogeneity, such as urban versus rural settings or localized outbreaks. For our simulation studies, we randomly allocated clusters to treatment arms across the study region while maintaining a 1:1 allocation ratio.

Although Section \ref{sec: design implications of spatial uncertainty} examines the potential design implications of cluster-level spatial predictive uncertainty, the current simulation study does not implement or evaluate adaptive recruitment or allocation strategies. Rather, treatment assignment remained fixed throughout all simulations, and our objective was to assess the performance of CRT-SMM under varying ICC and spatial dependence settings.

In real-world CRTs, especially within the National Cancer Institute’s Community Oncology Research Program (NCORP), estimated ICCs that are theoretically valid ranged from 0 to 0.50 \cite{snavelyIntraclusterCorrelationCoefficients2025}. Although many trials assume ICC to be less than or equal to 0.10 \cite{campbellClusterTrialsImplementation2001, adamsPatternsIntraclusterCorrelation2004, elleyIntraclassCorrelationCoefficients2005,murrayIntraclassCorrelationCommon1994, murrayIntraclassCorrelationMeasures1995}, \textcite{snavelyIntraclusterCorrelationCoefficients2025} highlighted the importance of assuming ICC to go beyond 0.10 if a robust calculation of sample size is to be made. 

To reflect this range of plausible real-world values, we assumed $\text{ICC} \in \{ 0.05, 0.15, 0.25 \}$ in our simulation study to cover both conservative and higher-end ICC scenarios in practice. We considered 6 different combinations of scenarios covering true treatment effect $\theta \in \{0-1.4 \ \text{by} \ 0.1\}$. Table \ref{tab:table1} shows scenarios we considered for each model. Note that for a chosen ICC, $\sigma_B^2$ and $\tau^2$ are calculated such that each explains 50\% of the variability attributable to $\sigma_B^2+\tau^2$. We considered a cluster size of $m=40$, a desired treatment effect of $\theta=0.6$, and $\sigma^2_W=2.25$. Although the required number of clusters to achieve 85\% power differed for each ICC, e.g., 17, 39, and 61 for each $\text{ICC} \in \{ 0.05, 0.15, 0.25 \}$, respectively, we fixed the number of clusters at 16 across all scenarios to isolate and evaluate the effects of spatial components independent of the number of clusters. The number of clusters in our simulations reflects real-world CRTs, where the median number of clusters has been reported as 21 in a random sample of 300 published trials \cite{taljaardInadequateReportingResearch2011}. 

We generated synthetic CRT datasets under the assumptions of a target treatment effect $\theta=0.6$, desired power of 0.85, and type I error rate of 0.05. Data generating mechanisms based on the specified fixed effects, variance components, and ICC values are outlined below: 

\begin{enumerate}
  \item[\text{Step 1.}] Specify fixed effects: We set the covariate effect $\gamma = 0.1$, the interaction effect $\delta = 0.1$. Covariates were simulated from $N(0,1)$ and marginal treatment effect simplifies to $\theta = \beta$, where we set treatment effect to vary from 0 to 1.4 in increments of 0.1.  
  \item[\text{Step 2.}] Define variance components: We fixed the individual-level variance $\sigma^2_W = 2.25$ to reflect that out target treatment effect has moderate effect size $\theta = 0.4 \sigma_W$. For each specified ICC, we partitioned variability attributable to $\sigma_B^2 +\tau^2$ using a proportion $f = \sigma_B^2/(\sigma_B^2 + \tau^2)$. The between-cluster variance $\sigma_B^2$ and spatial variance $\tau^2$ were computed as 
  $$
  \sigma_B^2 = \frac{\sigma_W^2}{\frac{1}{\text{ICC}} - \frac{1}{f}}, \  \tau^2 = \frac{(1-f) \sigma_B^2}{f},
  $$
where we set $f=0.5$ in our simulation, allocating equal variability to the cluster-level and spatial random effects. 
  \item[\text{Step 3.}] Incorporate spatial structure: We consider ICC values ICC $\in \{0.05, 0.15, 0.25\}$, which determine the corresponding $\sigma_B^2$ and $\tau^2$ values through variance partitioning in Step 2. The spatial decay parameter $\phi$ was set to either 1.5 or 3.5, chosen relative to the unit size of the simulated grid. 

  \item[\text{Step 4.}] Set sample size: 16 clusters (grid size $4 \times 4$) and $m=40$ individuals per cluster are considered. This setting corresponds to the target effect size as $\theta= 0.6$, desired power 0.85, with a two-sided type I error rate 0.05. The required sample size per arm to achieve this setting is calculated as 
  $$ N = 2 \frac{\sigma_W^2(Z_{1-\beta} + Z_{\alpha/2})^2}{\theta^2}\text{Deff}, $$
  where clustering is captured through design effect $\text{Deff} = 1 +(m-1) \cdot \text{ICC}$ \cite{campbellDevelopmentsClusterRandomized2007, arnupUnderstandingClusterRandomised2017}. The total number of clusters is then obtained as $2N/m$, given equal allocation to treatment and control arms. In a sample size calculation, design effect differs by the specific design (e.g., crossover, parallel, stepped-wedge) and study outcome types (e.g., continuous, binary). In our simulation, we use the specific case of parallel CRTs with a continuous outcome \cite{hemmingKeyConsiderationsDesigning2023}. 
  \item[\text{Step 5.}] Assign treatment: Clusters were allocated to treatment or control in a 1:1 ratio randomly. 
  \item[\text{Step 6.}] Generate outcomes: For each individual, outcomes were generated as a combination of treatment and covariate effects, cluster-level random effects, spatial random effects, and individual noise. We specifically simulated from a Gaussian random field with exponential covariance function. For simplicity, we considered one continuous covariate.
\end{enumerate}

Throughout our simulation, we utilized a novel informative Penalized Complexity (PC)-prior proposed by \textcite{simpsonPenalisingModelComponent2017}, as PC-priors are suited for constructing latent effects, invariant to reparameterization, connected to Jeffreys' priors, and robust regarding the choice of user defined parameter flexibility which is straightforward to apply through INLA. We chose weakly informative priors which aligns with standard practice for CRT analyses, where limited external information is available. In the Bayesian context, however, incorporating a well-justified informative prior can add value to statistical analysis and may be considered when appropriate \cite{jonesBayesianStatisticsDesign2021}. In our study, we utilized PC-priors on the standard deviation of the cluster-level and spatial variability, given the fact that the PC-prior is computationally efficient in fitting Bayesian hierarchical models in INLA \cite{simpsonPenalisingModelComponent2017}. The details specification of priors are given in the Section A2 of Supplementary Materials.

\subsection{Performance metrics}
The posterior probability of the treatment effect being greater than $\Delta$ can be approximated as 
\begin{align*}
    Pr(\hat{\theta} > \Delta|\bm{D}) \approx \frac{1}{M}\sum_{r=1}^M 1[P_{H_0}(\hat{\theta}^{(r)} > \Delta]
\end{align*}
where $\bm{D}$ is data accumulated at the end of the trial, $M$ is the number of posterior samples drawn from marginal treatment effects, and $\hat{\theta}^{(r)}$ is a $r^{\text{th}}$ posterior sample. We used INLA, a method for approximate Bayesian inference, where the marginal posterior distribution $ Pr(\hat{\theta} > \Delta|\bm{D})$ is approximated numerically rather than averaged from MCMC samples. 

For each scenario, $S=10,000$ number of trials were simulated, and depending on the decision boundary $1-\alpha$, overall false positive rate (FPR) and power are calculated. In our simulation, we chose $\Delta =0$ for simplicity. The overall FPR is computed as the proportion of simulations falsely declaring meaningful efficacy, where falsely rejecting $H_0$ is denoted as $P_{\theta = 0}(\hat{\theta}>\Delta|\bm{D}) > 1-\alpha$. The overall power is computed as the proportion of simulations correctly identifying the meaningful efficacy, where correctly rejecting $H_0$ is denoted as $P_{\theta>0}(\hat{\theta}>\Delta|\bm{D}) > 1-\alpha$.

The percent relative error (\%RE), which is an informative performance measure that represents under or over-prediction \cite{katValidationMetricBased2012}, is calculated as 
\begin{align*}
    \text{\%RE} = \left( \frac{\text{modSE}}{\text{empSE}} -1 \right) \times 100 \%,
\end{align*}
where modSE refers to the model‑derived posterior standard error obtained from each simulation, while empSE denotes the empirical standard error calculated from the distribution of mean estimates across repeated simulations. A large positive \%RE indicates that the estimate intervals are too wide (i.e., conservative), whereas a large negative \%RE indicates the intervals are too narrow (i.e., anti-conservative). When the uncertainty of a parameter is underestimated, the power and type I error rate become inflated because it becomes easier to reject the null hypothesis. Also, this leads to under-coverage of the true parameter value.

Bias and mean squared error (MSE) were used to evaluate estimator performance across repeated simulations. Coverage probability was used to assess the posterior uncertainties. For each simulation replicate, 95\% credible interval for the treatment effect $\theta$ was constructed.

\subsection{Comparison methods}\label{sec:comparison methods}
We considered a set of models to evaluate the impact of incorporating spatial effects in clinical trials. We compared CRT-SMM with four nonspatial methods applied to simulated CRTs. (i) na\"{\i}ve fixed effects model (CRT-FM-na\"{\i}ve), which ignores the cluster effect entirely, (ii) fixed effects model (CRT-FM), which accounts for the cluster effect as a fixed term, (iii) mixed effects model (CRT-MM), which accounts for the cluster effect as a random term, and (iv) cluster-level analysis (CRT-cluster). Across individual-level models (i)-(iii), we used the same individual-level mean structure $\mu_{ij} =  \beta z_{i} + \bm{x}_{ij}'\bm{\gamma} + z_{i} \cdot \bm{x}_{ij}' \bm{\delta}$ as defined in equation (\ref{eq:mu}), and included individual-level variation $\epsilon_{ij} \sim N(0, \sigma_W^2)$. For cluster-level model (iv), we used the averaged mean structure $\bar{\mu}_i = \beta z_i + \bar{\bm{x}_i}'\gamma + z_i\bar{\bm{x}_i}'\bm{\delta}$.

\textit{\textbf{CRT-FM-na\"{\i}ve}}\\
First, we consider the CRT na\"{\i}ve fixed effects model (CRT-FM-na\"{\i}ve)
\begin{align*}
    Y_{ij}|\beta, \bm{\gamma}, \bm{\delta}, \sigma_W^2 \sim N(\mu_{ij}, \ \sigma_W^2),
\end{align*}
which ignores cluster-level variability entirely. This ``complete pooling" approach is known to underestimate uncertainty, resulting in spuriously inflated statistical power and type I error \cite{agrestiCategoricalDataAnalysis2011, bartelsFIXEDRANDOMEFFECTS}. We included this model as a benchmark despite its anticipated inferential shortcomings.

\textit{\textbf{CRT-FM}}\\
Next, we considered the CRT fixed-effects model (CRT-FM), where dummy cluster membership indicators with unknown fixed effects are specified to absorb all the between-cluster variations \cite{bartelsFIXEDRANDOMEFFECTS}, formulated as follows:
\begin{align*} 
    Y_{ij}|\beta, \bm{\gamma}, \bm{\delta}, \sigma_W^2 \sim N(\mu_{ij}+ \xi_{g(i)}, \ \sigma_W^2),
\end{align*}
where $\xi_{g(i)}=\sum_{k=2}^{I}\xi_k \cdot \mathds{I}\{g(i) = k \}$ denotes the fixed effect for cluster $g(i) \in \{2,..., I\}$, represented through cluster indicator $\mathds{I}\{ \cdot \}$ with first cluster serving as the reference group, i.e., $\xi_{g(1)}=0$ \cite{tutzModelingClusteredHeterogeneity2014}. In parallel CRTs, treatment is assigned at the cluster level, resulting in perfect multicollinearity between treatment and cluster indicators, particularly under ITT framework. To improve identifiability, we assigned moderately informative priors $N(0,1)$ to all fixed effects.

\textit{\textbf{CRT-MM}}\\
We consider the CRT mixed-effects model (CRT-MM) that accounts for the unobserved heterogeneity by incorporating cluster-level random effects, or random intercept, $u_{i}$. The formula for CRT-MM is as follows: 
\begin{align*}
   Y_{ij}|\beta, \bm{\gamma}, \bm{\delta}, \bm{u}, \sigma_W^2  \sim N(\mu_{ij} + u_i, \  \sigma_W^2).
\end{align*}

\textit{\textbf{CRT-Cluster}}\\
We also compared our models analyzed with cluster-level summary measures rather than individual-level to provide the ground for simplicity. The CRT-cluster is
\begin{align*}
    \bar{Y}_{i}|\beta, \bm{\gamma}, \bm{\delta}, \sigma_{W}^2 \sim N ( \bar{\mu}_{i}, \ \sigma_{W}^2),
\end{align*} 
where $\bar{\mu}_i = \beta z_i + \bar{\bm{x}_i}'\gamma + z_i\bar{\bm{x}_i}'\bm{\delta}$. In a cluster-level analysis, cluster-level covariates can be used directly instead of aggregating individual-level covariates $\bm{x}_{ij}$ to $\bar{\bm{x}_i}$. For consistency across methods, we averaged individual-level covariates within each cluster to obtain $\bar{\bm{x}_i}$ in the simulation study.

\subsection{Simulation results}\label{sec:results}
\begin{figure}[p]
        \centering
        \includegraphics[width=0.82\linewidth]{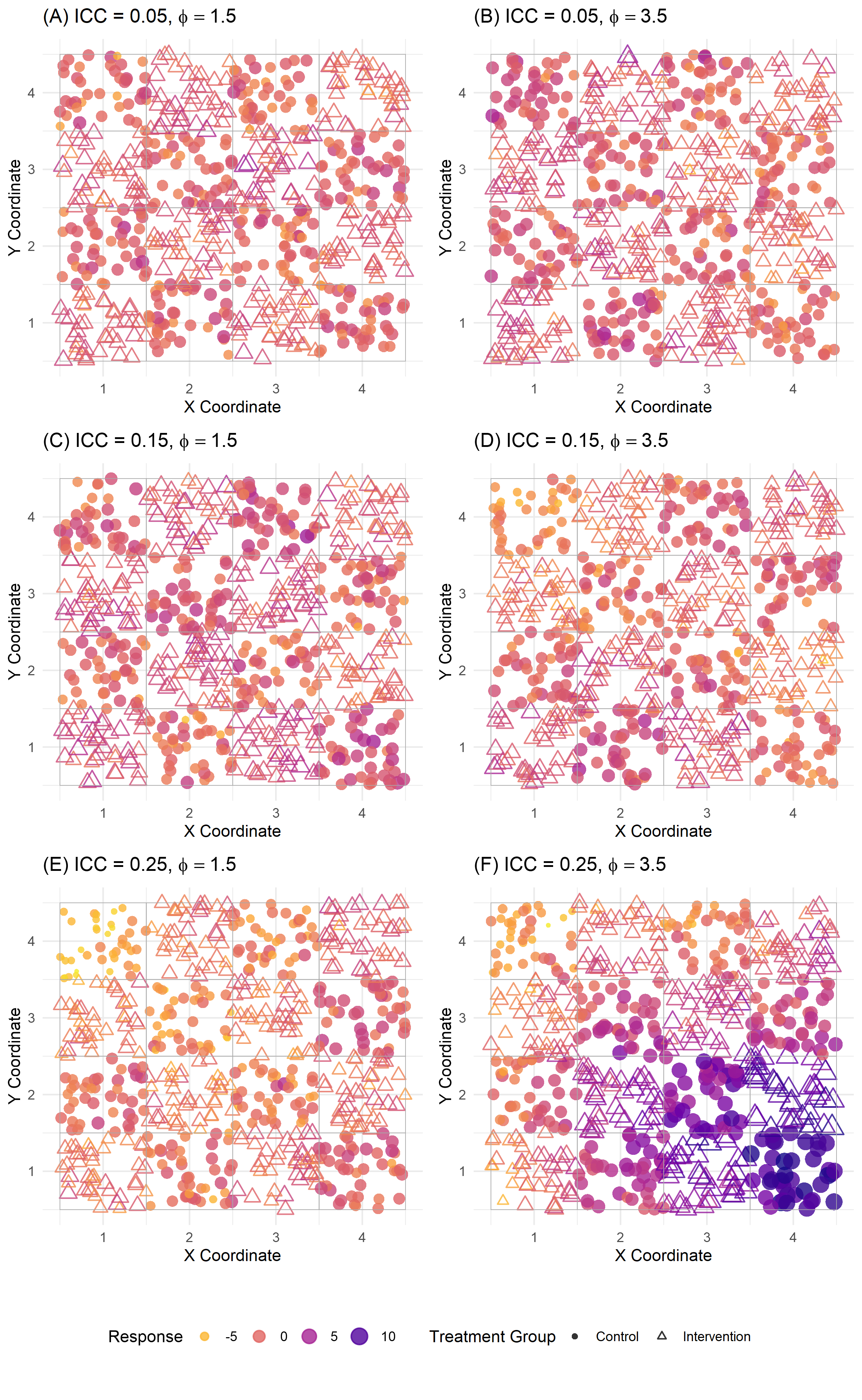}
        \caption{Patient responses from a single simulation replicate with restricted checkerboard structure across scenarios when true $\theta = 0.5$.}
        \label{fig:fig1}
\end{figure}

A total of 10,000 simulations were conducted for each scenario using R (version 4.3.3) \cite{R-base}. Figure \ref{fig:fig1} is a single simulation replicate across scenarios when true treatment effect is $\theta = 0.5$, under a checkerboard restricted randomization of clusters. Although our simulation assigns treatments randomly, we present an example with a checkerboard structure to better illustrate the distribution of individual responses across treatment groups. The individual responses are generated on a $4 \times 4$ grid, where the correlation between the centers of the farthest diagonal (from the upper left to the lower right) is given by $\exp(-{3\sqrt{2}}/{\phi})$. For example, Scenarios A, C, E have a correlation of $\exp({-3 \sqrt{2}}/{1.5}) = 0.059$, and Scenarios B, D, F have a correlation of $\exp({-3 \sqrt{2}}/{3.5}) = 0.297$ at the maximum grid-center distance. The figure demonstrates that within each treatment group, individuals in close proximity tend to have similar response values, whereas those farther apart show greater differences. This spatial dependency becomes more prominent as the ICC and spatial range $\phi$ increase. 

Table \ref{tab:table2} summarizes 10,000 simulation posterior mean estimates $\hat{\theta}$ and standard deviation $\hat{\sigma_{\theta}}$ for true treatment values $\theta \in \{0,0.3,0.6 \}$ under Scenarios A-F. When $\theta=0$, all methods produced unbiased estimates, but their variance behavior differed. CRT-FM-na\"{\i}ve consistently underestimated uncertainty, with very small standard deviations in Scenarios A–C and only modest increases in Scenarios D–F. CRT-FM provided unbiased estimates with somewhat larger standard deviations, partially correcting the underestimation. CRT-MM and CRT-cluster also produced unbiased means, but their variances escalated sharply with stronger spatial correlation, reaching as high as 5.69 and 6.33 in Scenario F. In contrast, CRT-SMM achieved unbiased estimates with moderate and stable variances (0.24–0.96) across all scenarios. 

When $\theta=0.3$ or $\theta = 0.6$, CRT-FM-na\"{\i}ve remained relatively unbiased but underestimated variance. CRT-FM, CRT-MM, and CRT-cluster showed relatively biased estimates and became severely biased under Scenario F, with CRT-MM and CRT-cluster exhibiting high uncertainty ranging 1.83-5.69 and 2.00-6.34, respectively, in Scenarios D-F. Meanwhile, CRT-SMM produced unbiased estimates with well-calibrated and stable variance across scenarios.

\begin{figure}[p]
    \centering
    \includegraphics[width=1\linewidth]{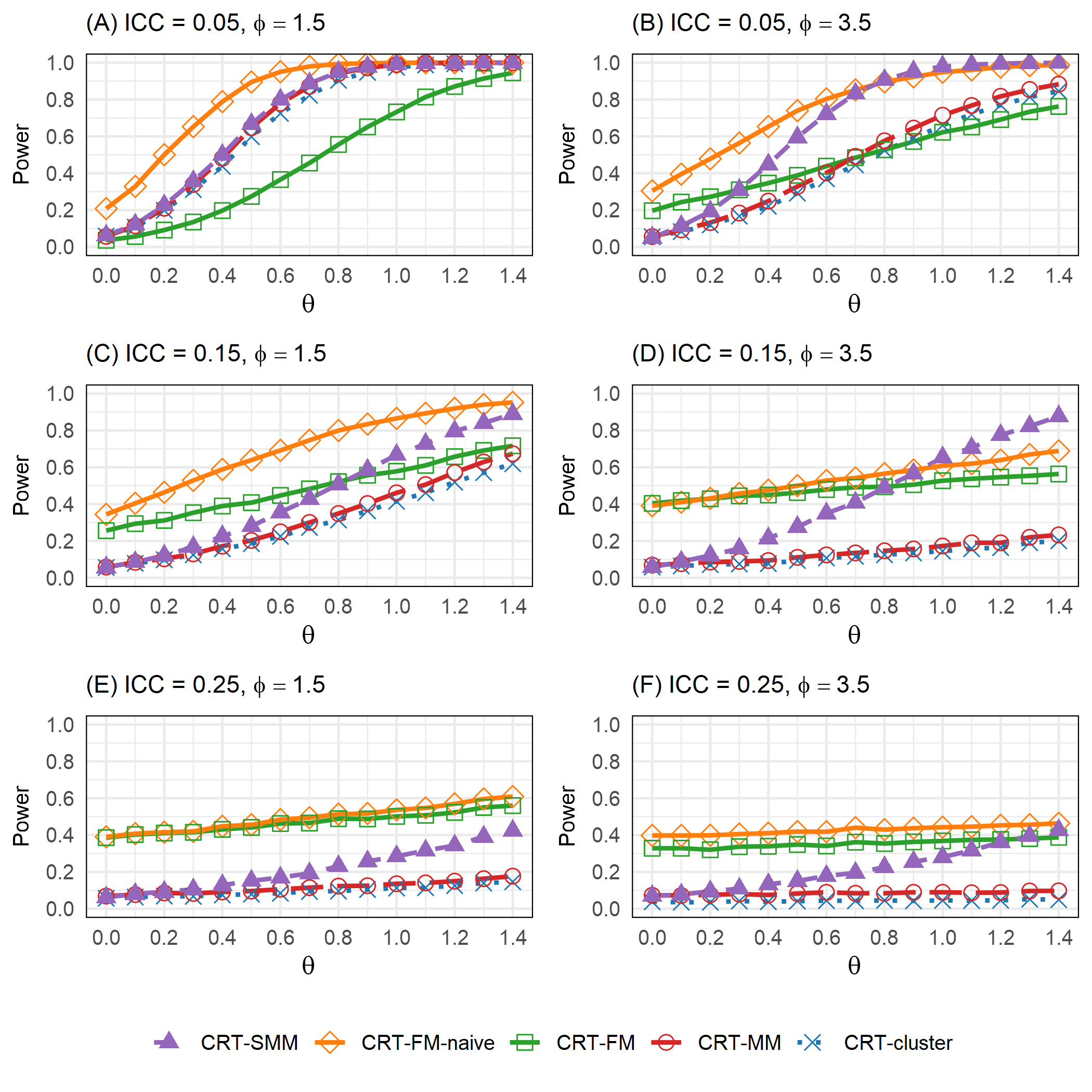}
    \caption{Power curves of 10,000 simulations across true treatment $\theta$'s presented by data generation scenarios. The number of clusters was fixed at 16, as determined by the CRT design calculation corresponding to ICC = 0.05.}
    \label{fig:fig2}
\end{figure}

Figure \ref{fig:fig2} displays power curves from 10,000 simulations across true treatment effects $\theta$ under Scenarios A-F. CRT-FM-na\"{\i}ve generally achieved the highest power across scenarios, although CRT-SMM surpassed it in Scenarios B and D as $\theta$ increased. 
CRT-FM performed worst in Scenario A but maintained moderate power in Scenarios B–F. CRT-MM achieved higher power when the spatial range parameter was $\phi=1.5$ compared to $\phi=3.5$ but its performance was substantially lower than that of CRT-SMM, unless ICC was low as 0.05. CRT-cluster followed a trend similar to CRT-MM. Our proposed method, CRT-SMM, approached the highest power levels when $\phi = 3.5$. It is important to note that the trial design fixed the number of clusters at 16 across all scenarios, slightly fewer than the 17 clusters suggested by sample size calculations to achieve 85\% power and a type I error rate of 0.05 when the target treatment effect is $\theta=0.6$, under ICC = 0.05. Although CRT-SMM did not surpass other methods in power at $\phi=1.5$, its upward trend suggested that it would eventually outperform other methods once the number of clusters is optimized for higher ICC values. 

\begin{figure}[p]
    \centering
    \includegraphics[width=1\linewidth]{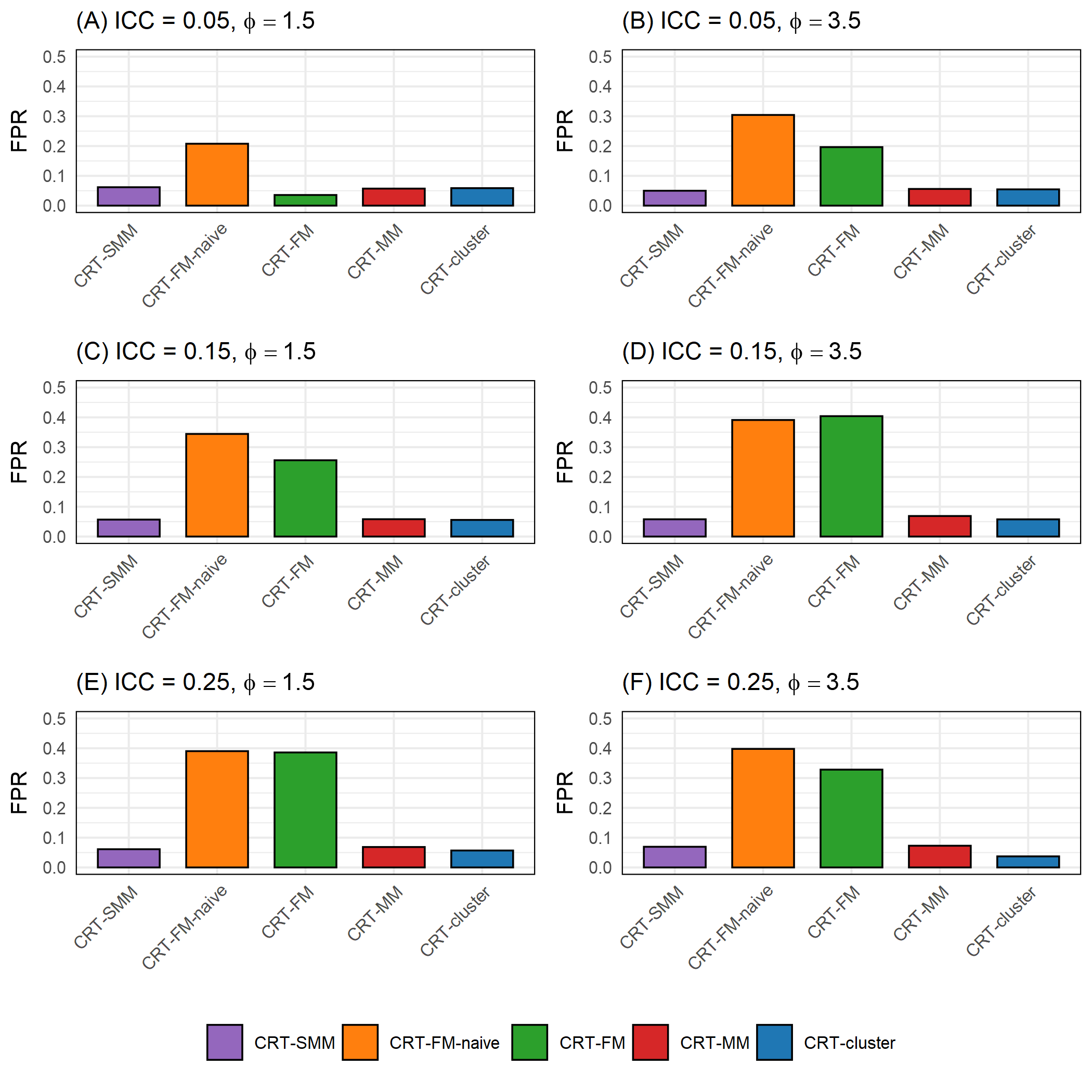}
    \caption{False positive rates (FPRs) of 10,000 simulations data generation scenarios.}
    \label{fig:fig3}
\end{figure}

The false positive rates in Figure \ref{fig:fig3} show that CRT-FM-na\"{\i}ve had the highest FPR ranging from 0.2 to 0.4, which represents an unacceptably inflated type I error rate. CRT-FM achieved an acceptable FPR close to 0.05 only in Scenario A, but exhibited similarly inflated rates as CRT-FM-na\"{\i}ve in Scenarios B-F. In contrast, CRT-MM, CRT-SMM, and CRT-cluster maintained acceptable FPRs across all scenarios (A–F). 

\begin{figure}[p]
    \centering
    \includegraphics[width=1\linewidth]{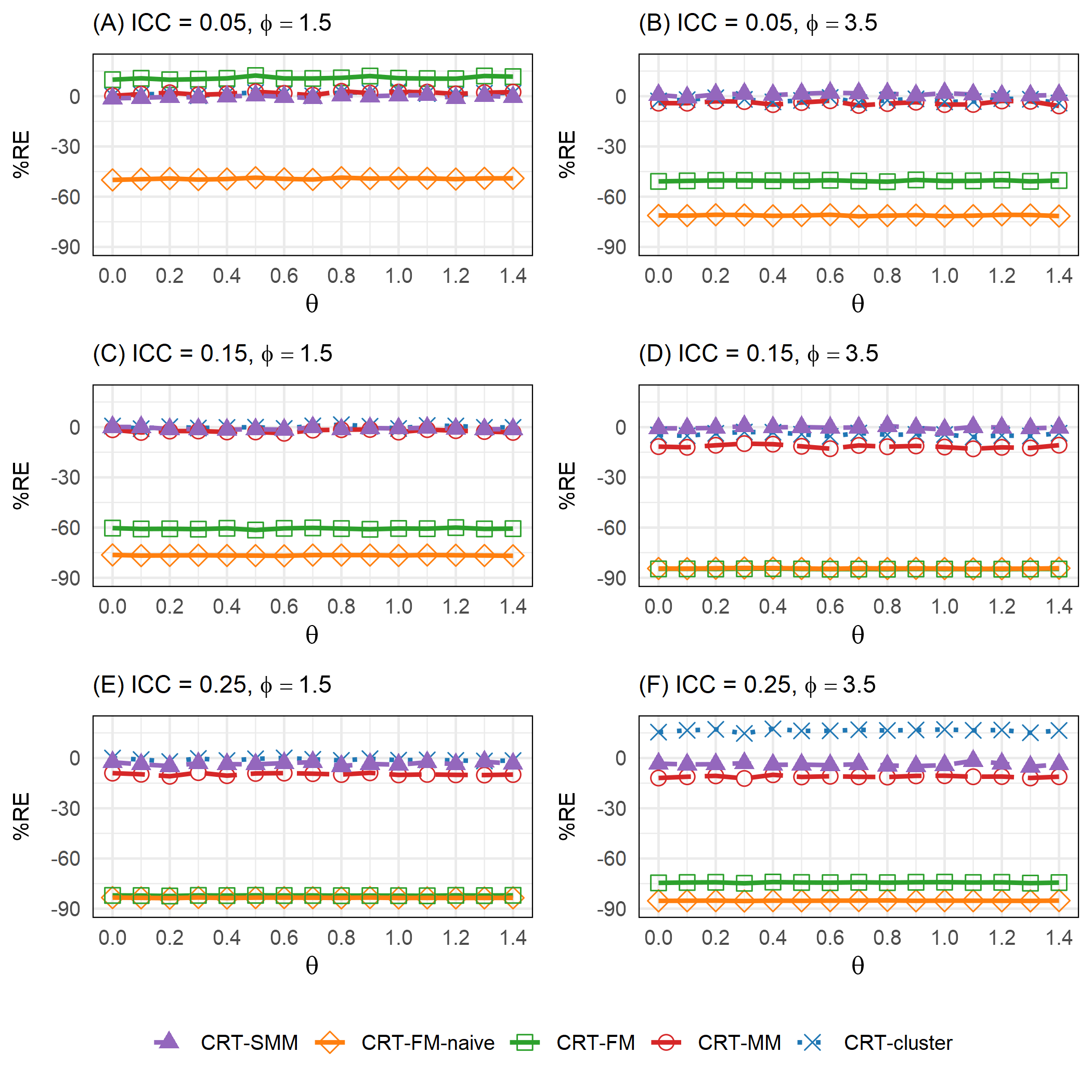}
    \caption{Percent relative error (\%RE) of 10,000 simulations across true treatment $\theta$'s presented by data generation scenarios.}
    \label{fig:fig4}
\end{figure}

Percent relative error (\%RE) provides a measure of the inaccuracy of estimates relative to the true value. Figure \ref{fig:fig4} displays \%RE from 10,000 simulations across all scenarios. CRT-FM-na\"{\i}ve exhibited the largest negative values, ranging from -45 to -90, indicating anti-conservative confidence intervals that led to inflated power, consistent with the power curves in Figure \ref{fig:fig2}. CRT-FM showed slightly positive values in Scenario A but similarly large negative values (–45 to –90) in Scenarios B–F, reflecting the inflated power curves in Figure \ref{fig:fig2}, except in Scenario A where it yielded the lowest power across values of $\theta$'s. CRT-MM produced near-zero \%RE in Scenarios A-C, with slightly negative values in Scenarios D–F. CRT-cluster showed near-zero \%RE in Scenarios A-E but has a positive value of approximately 15 in Scenario F. Importantly, CRT-SMM maintained \%RE values consistently near zero across all scenarios (A–F), demonstrating the most reliable and well-calibrated performance among all methods.

Additional figures (Figure S2, S3, S4) present the coverage probability of $\theta$, bias, and MSE of 10,000 simulations across all scenarios. The results show that CRT-MM, CRT-cluster, and CRT-SMM consistently achieved coverage probabilities close to 1 in all scenarios (A–F), whereas CRT-FM-na\"{\i}ve and CRT-FM exhibited substantially lower coverage, with the exception of CRT-FM attaining the highest coverage in Scenario A. The complete pooling approach (CRT-FM-na\"{\i}ve) in Figure S3 demonstrates that failure to account for unobserved heterogeneity induces additional bias \cite{hsiaoAnalysisPanelData2022, skrondalGeneralizedLatentVariable2004}, which seemed to manifest when ICC is high or when the spatial range is large. 

In summary, our proposed spatial approach (CRT-SMM) generally outperformed methods that ignore spatial variability spatial correlation was long-ranged ($\phi$=3.5) and remained competitive under shorter-range spatial correlation ($\phi=1.5$). In addition, FPR assessed type I error control, \%RE evaluated variance calibration, coverage probability examined uncertainty validity, and bias and MSE measured estimator accuracy and overall quality. Together, these metrics demonstrated that CRT-SMM excelled in stable and reliable inference.

Sensitivity analyses for different covariance functions and priors for the Gaussian process in CRT-SMM are provided in Section A6 of Supplementary Materials. Table S1 outlines six prior settings based on varying covariance functions and hyperpriors for the spatial parameters $\phi$ and $\tau^2$. For these analyses, we focus on CRT-SMM under the setting with true ICC = 0.05 (corresponding to $\tau^2 = 0.125$) and $\phi=3.5$. Table S2 summarizes results from 10,000 simulations when the true treatment effects were set as $\theta \in \{0, 0.3, 0.6 \}$. The findings demonstrate that CRT-SMM remains robust across different prior specifications.

\section{Application} \label{sec:data applcation}
\subsection{Data description}
To demonstrate the practical implementation of the proposed framework, we used publicly available 2015 Demographic and Health Surveys (DHS) (https://dhsprogram.com/) data from \textcite{ZIMSTAT2016DHS}, which provided georeferenced clusters (i.e., groupings of households). To protect confidentiality, only randomly displaced GPS locations of clusters were available. Although the survey data did not arise from a CRT and should be interpreted as observational, they retain a clustered sampling design and spatial structure, suitable for demonstrating our proposed CRT-SMM approach.  The effect estimates in this example should be interpreted as associations rather than causal effects. 

Insecticide-treated bed nets (ITNs) are known to reduce child's death and illness from malaria \cite{lengelerInsecticidetreatedBedNets2004, takkenInsecticidetreatedBednetsHave2002, lindbladeCohortStudyEffectiveness2015}. We investigated the estimated effect of ITNs on whether children under age 5 who slept under ITNs had reduced reports of severe symptoms (illness accompanied by fever and cough) in the past two weeks. The dataset included a total of 398 clusters, with cluster size ranging from 2 to 34 (median 12), comprising 657 treated and 4,544 control observations. In this setting, children's usage of ITNs is collected at the household level and treatment status (i.e., ITN treated vs comparison) varied within clusters, differing from typical geographic CRTs.

\begin{figure}[h]
    \centering
    \includegraphics[width=0.9\linewidth]{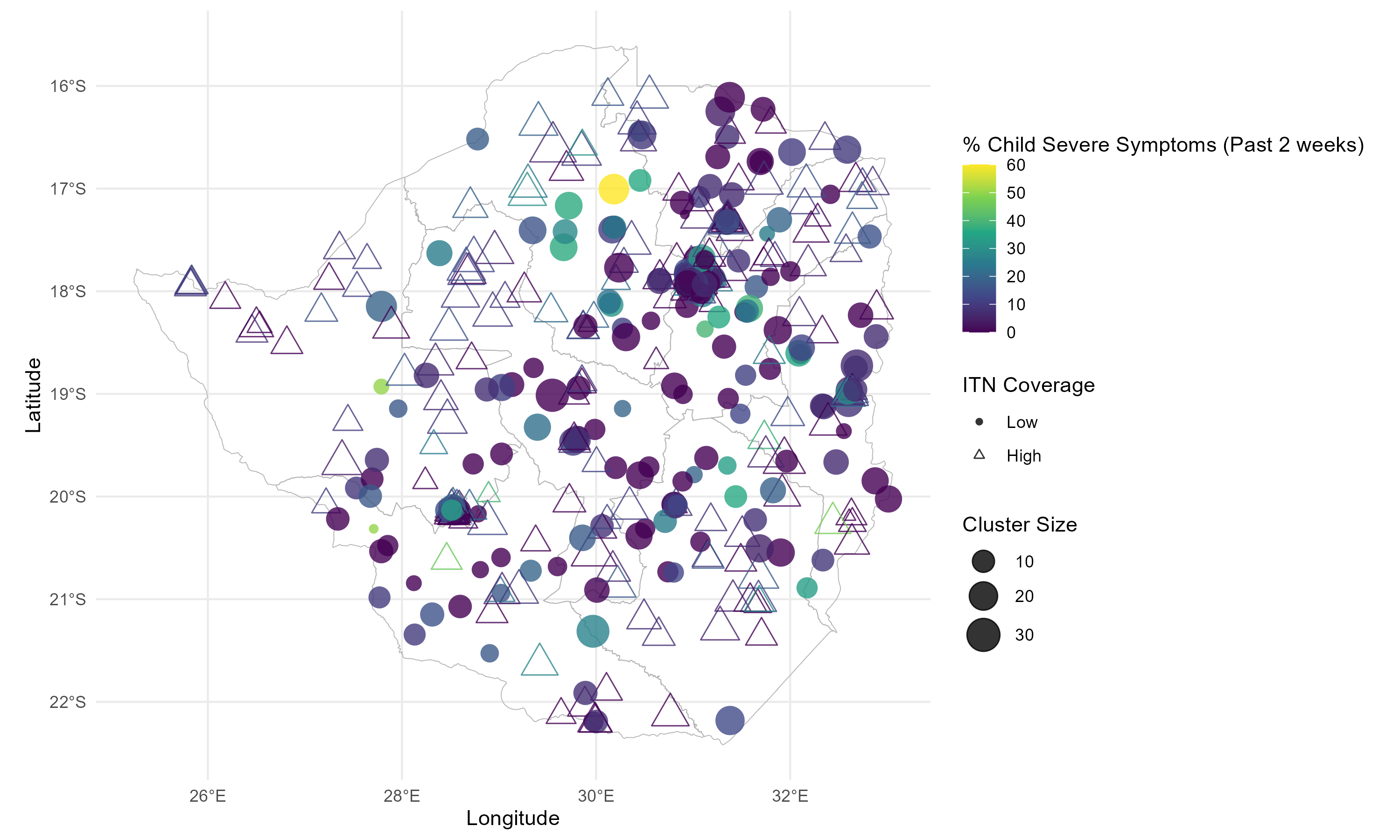}
    \caption{DHS clusters in Zimbabwe \cite{ZIMSTAT2016DHS}. Each point represents a cluster with summarized data on ITN coverage and the percentage of children reporting severe symptoms.}
    \label{fig:DHS}
\end{figure}

Figure \ref{fig:DHS} shows the cluster points in Zimbabwe where different number of households participated in the survey. Each cluster includes multiple children with information on ITN use and report of severe symptoms experienced in the past two weeks. Because household-level GPS coordinates were unavailable, we summarized the data at the cluster level by computing ITN coverage (low: below median; high: above median) and the percentage of children reporting severe symptoms.

As the outcomes were binary, we extended equation (\ref{eq:y_s}) to a generalized linear mixed-effects model using logit link:  
\begin{align*}
Y_{ij} &\sim \text{Bernoulli}(p_{ij}),\\
    \text{logit}(p_{ij}) &= \beta z_{ij} + \bm{x}_{ij}'\bm{\gamma} + u_{i} + w(\bm{s}_{i}),
\end{align*}
where $Y_{ij}$ is severe symptoms in the past two weeks (Yes/No), $p_{ij}$ is the probability that child $j$ in cluster $i$ experienced severe symptoms and $\bm{x}_{ij}$ includes child's age and household wealth index. The CRT-SMM was compared to CRT-FM-na\"{\i}ve, CRT-FM, and CRT-MM. Note that CRT-cluster was not applicable in our analysis, as clustered observational studies contained both treated and untreated individuals within the same cluster.

\subsection{Application results}
The global Moran's I on resi duals of fitted logistic regression showed that the spatial heterogeneity existed (p-value < 0.05). We also plotted an empirical variogram on the residuals to evaluate how much data pairs differed by distance, which showed a noisy pattern with a large nugget effect, indicating that nearby points are already dissimilar \cite{pitardExplorationNuggetEffect1994}. This may be partly attributable to the randomly displaced data to protect the confidentiality and the to large sampling interval exceeding the underlying correlation range \cite{oliverTutorialGuideGeostatistics2014a}. However, specifying a spatial range parameter with plausible short distance may still improve model stability and inference.

The PC-prior was calibrated to reflect the spatial scale of the study region in Zimbabwe. Specifically, we specified weakly informative priors for the spatial range, guided by the empirical variogram presented in Figure S5 in the Supplementary Materials. Because the DHS data are measured numerically in meters, unlike the interval 4-by-4 scale, the prior for the spatial range was adjusted accordingly, with $P(\phi<50,000)=0.5$ (the maximum distance between the points was 836,917m). All other priors are specified as weakly informative. Sensitivity analyses on wider spatial range priors yielded no meaningful changes in results.

Table \ref{tab:dhs results} shows the result of effect estimates and model fit for ITN usage on severe symptoms in children under 5. The odds ratios (OR) were consistent across all specifications and close to the null value (OR $\approx$ 1), with 95\% credible intervals (CrI) including 1, suggesting no strong evidence of an association between ITN use and severe symptoms after adjusting for child’s age and household wealth index. Accounting for clustering substantially improved model fit, with both CRT-MM and CRT-SMM showing markedly lower Deviance Information Criterion (DIC) and Watanabe-Akaike Information Criterion (WAIC) compared to the fixed-effects models, indicating important cluster-level heterogeneity. The ICC from CRT-MM was 0.147 (95\% CrI: (0.108, 0.188)), suggesting moderate within-cluster correlation.

Including a spatial random effect (CRT-SMM) yielded a modest additional improvement in fit and reduced the ICC to 0.107 (95\% CrI: (0.011, 0.174)), implying that some minor cluster-level variability is explained by spatial structure. Despite the weak and noisy spatial pattern, likely due to random displacement of cluster coordinates for confidentiality and large distances between cluster points, CRT-SMM slightly improved the model fit, indicating that even weak spatial dependence can help explain underlying heterogeneity.  

\section{Discussion}\label{sec:discussion}
We have provided comprehensive methods for CRTs under varying spatial correlations and ICC, highlighting substantial differences in power, FPR, \%RE, coverage probability, bias, and MSE. Our findings demonstrate that methods that fail to demonstrate spatial dependency consistently underestimate uncertainty and inflate type I errors. These deficiencies manifest in scenarios where ICC is as high as 0.15 or 0.25 or when spatial range $\phi$, also referred to as the decay parameter that controls the reach of the spatial distance (in Supplementary Materials) is as high as 3.5. This spatial range is reasonable for a $4 \times 4$ grid where in a real-world, shared environmental or health system factors could impact each individual's response depending on the trial tasks and should be considered through spatial dependencies. 

 Our model which incorporates individuals' locations to account for spatial heterogeneity offers several benefits. First, it captures geographic variation in outcomes, recognizing that the effect of an intervention on health outcomes may vary spatially due to environmental, socioeconomic, and healthcare access factors. Second, it accounts for spillover effects; for example, in a vaccine trial, herd immunity can arise in groups that received the intervention. Failing to account for spatial dependency may therefore lead to incorrect estimates of treatment effects. Third, it improves efficiency in sample size calculation as understanding spatial correlation allows for more informed cluster size selection, reducing the number of participants needed. Lastly, by quantifying cluster-level spatial predictive uncertainty, the CRT-SMM framework provides information that can be leveraged for future cluster recruitment strategies.

The theoretical derivation and empirical illustration of spatial predictive uncertainty inform the potential design implications for treatment allocation and recruitment strategies that increase estimation efficiency. In particular, clusters with larger $\text{APV}_i$ inflate the variance of estimates through their contribution to the covariance structure. This observation suggests a natural extension to a two-stage adaptive design, where $\text{APV}_i$ can be computed after the first stage (e.g., enrolling half of the planned sample size), and then allocating the remaining participants to clusters with higher predictive uncertainty in the second stage. An uncertainty-informed allocation strategy would remain appropriate unless a major imbalance (e.g., 80\% of participants belonging to only 20\% of clusters) arises following uncertainty-driven allocation, as moderate cluster size inequality generally has little impact on statistical power \cite{guittetPlanningClusterRandomized2006}.

Our methods can be extended to stepped-wedge designs, in which more clusters are exposed to the intervention toward the end of the study than at its early stage; to crossover designs, in which clusters are randomized sequentially to two or more arms and eventually each cluster receives both arms, serving as its own control for treatment comparisons \cite{lewisStatisticalPrinciplesClinical1999}; and to randomized or observational studies with geocoded data, where treatment varies at the individual level but spatial and cluster-level dependence remains.

Key considerations for designing and analyzing CRTs must be carefully addressed. Compared with individually randomized trials, CRTs typically require larger sample sizes, involve added complexities, and have a greater risk of bias \cite{hemmingKeyConsiderationsDesigning2023}. Nevertheless, despite the belief that CRTs lack precision, Raudenbush (1997) \cite{raudenbushStatisticalAnalysisOptimal} argued that efficient modeling with the use of information at each level, combined with careful choice of covariates and sound planning of the design, can significantly improve the precision of CRTs. Building on this, our demonstration of geostatistical methods for geographical CRTs highlights the potential for broader applications by incorporating spatial methods and integrating not only individual covariates but also neighborhood, socioeconomic, and environmental factors into the analytic framework.

\begin{table}[h]
\centering
\begin{threeparttable}
\caption{Scenarios for data generation with a fixed cluster size of $m=40$, desired treatment effect $\theta=0.6$, and $\sigma^2_W=2.25$, using 16 clusters with a 1:1 allocation ratio.}
\label{tab:table1}
\begin{tabularx}{\textwidth}{@{}>{\centering\arraybackslash}X|
  >{\centering\arraybackslash}X
  >{\centering\arraybackslash}X
  >{\centering\arraybackslash}X
  >{\centering\arraybackslash}p{3cm}@{}}
\toprule
Scenario & ICC & $\sigma_B^2$ & $\tau^2$ & $\phi$ \\
\midrule
A & 0.05 & 0.125 & 0.125 & 1.5  \\
B & 0.05 & 0.125 & 0.125 & 3.5  \\
C & 0.15 & 0.482 & 0.482 & 1.5 \\
D & 0.15 & 0.482 & 0.482 & 3.5  \\
E & 0.25 & 1.125 & 1.125 & 1.5  \\
F & 0.25 & 1.125 & 1.125 & 3.5  \\
\bottomrule
\end{tabularx}
\begin{tablenotes}
\footnotesize
 \item[]Note: For the spatial range parameter $\phi$, smaller values correspond to faster decay with distance, meaning that points further apart become much less correlated; For spatial variability $\tau^2$, a larger value means higher spatial heterogeneity, which implies stronger local effects.
\end{tablenotes}
\end{threeparttable}
\end{table}

\begin{table}[h] 
\centering
\begin{threeparttable}
\caption{Posterior mean ($\hat{\theta}$) and standard deviation ($\hat{\sigma}_{\theta}$) of the treatment effect from 10,000 simulations for true values $\theta=0, 0.3$, and $0.6$ across scenarios.}
\label{tab:table2}

\begin{tabularx}{\textwidth}{@{}
  >{\centering\arraybackslash}p{1cm}@{}|
  >{\centering\arraybackslash}p{1.6cm}@{}|
  >{\centering\arraybackslash}X
  >{\centering\arraybackslash}X
  >{\centering\arraybackslash}X
  >{\centering\arraybackslash}X
  >{\centering\arraybackslash}X
  >{\centering\arraybackslash}X
  >{\centering\arraybackslash}X
  >{\centering\arraybackslash}X
  >{\centering\arraybackslash}X
  >{\centering\arraybackslash}X}
\toprule
 &
 & \multicolumn{2}{c}{CRT-SMM} 
 & \multicolumn{2}{c}{CRT-FM-na\"{\i}ve} 
 & \multicolumn{2}{c}{CRT-FM} 
 & \multicolumn{2}{c}{CRT-MM} 
 & \multicolumn{2}{c}{CRT-cluster}\\
\cmidrule(lr){3-4} \cmidrule(lr){5-6} \cmidrule(lr){7-8} \cmidrule(lr){9-10} \cmidrule(lr){11-12}  
$\theta$ & Scenario & $\hat{\theta}$ & $\hat{\sigma}_{\theta}$ 
& $\hat{\theta}$ & $\hat{\sigma}_{\theta}$
& $\hat{\theta}$ & $\hat{\sigma}_{\theta}$ 
& $\hat{\theta}$ & $\hat{\sigma}_{\theta}$
& $\hat{\theta}$ & $\hat{\sigma}_{\theta}$\\
\midrule
  & A & 0.00 & 0.24 & 0.00 & 0.12 & -0.00 & 0.38 & 0.00 & 0.25 & 0.00 & 0.27 \\
  & B & -0.00 & 0.27 & 0.00 & 0.14 & -0.00 & 0.39 & 0.00 & 0.46 & 0.00 & 0.50 \\
0 & C & 0.00 & 0.48 & 0.00 & 0.16 & 0.01 & 0.39 & 0.00 & 0.66 & 0.01 & 0.72 \\
  & D & -0.00 & 0.49 & 0.02 & 0.34 & 0.03 & 0.41 & 0.02 & 1.84 & 0.02 & 2.01 \\
  & E & -0.02 & 0.95 & -0.04 & 0.46 & -0.03 & 0.46 & -0.04 & 2.36 & -0.04 & 2.58\\
  & F & 0.02 & 0.96 & 0.04 & 1.53 & -0.01 & 0.79 & 0.01 & 5.69 & -0.00 & 6.33\\

  \midrule
    & A & 0.30 & 0.24 & 0.30 & 0.12 & 0.26 & 0.38 & 0.30 & 0.25 & 0.30 & 0.27 \\
    & B & 0.30 & 0.27 & 0.30 & 0.14 & 0.25 & 0.39 & 0.30 & 0.46 & 0.29 & 0.50 \\
0.3 & C & 0.30 & 0.48 & 0.30 & 0.16 & 0.26 & 0.39 & 0.30 & 0.66 & 0.30 & 0.72 \\
    & D & 0.31 & 0.49 & 0.27 & 0.34 & 0.23 & 0.41 & 0.26 & 1.83 & 0.27 & 2.00 \\
    & E & 0.30 & 0.96 & 0.31 & 0.46 & 0.25 & 0.46 & 0.29 & 2.36 & 0.29 & 2.58 \\
    & F & 0.31 & 0.96 & 0.31 & 1.53 & 0.10 & 0.79 & 0.21 & 5.67 & 0.17 & 6.32 \\

  \midrule
    & A & 0.60 & 0.24 & 0.60 & 0.12 & 0.51 & 0.38 & 0.60 & 0.25 & 0.60 & 0.27 \\
    & B & 0.60 & 0.27 & 0.60 & 0.14 & 0.51 & 0.39 & 0.60 & 0.46 & 0.60 & 0.50 \\
0.6 & C & 0.60 & 0.48 & 0.60 & 0.16 & 0.51 & 0.39 & 0.59 & 0.66 & 0.59 & 0.72 \\
    & D & 0.60 & 0.49 & 0.62 & 0.34 & 0.54 & 0.41 & 0.59 & 1.84 & 0.59 & 2.01 \\
    & E & 0.60 & 0.96 & 0.62 & 0.46 & 0.48 & 0.46 & 0.58 & 2.36 & 0.58 & 2.59 \\
    & F & 0.60 & 0.96 & 0.54 & 1.53 & 0.19 & 0.8 & 0.37 & 5.69 & 0.32 & 6.34 \\
  
\bottomrule
\end{tabularx}
\begin{tablenotes}
\footnotesize
 \item[]
\end{tablenotes}
\end{threeparttable}
\end{table}

\begin{table}[h]
\centering
\begin{threeparttable}
\caption{Comparison of ITN usage effect estimates on severe symptoms. ICC and model fit statistics across models are presented.}
\label{tab:dhs results}

\begin{tabularx}{\textwidth}{@{}
  >{\centering\arraybackslash}p{2.8cm}|
  >{\centering\arraybackslash}p{4cm}
  >{\centering\arraybackslash}p{4cm}
  >{\centering\arraybackslash}X
  >{\centering\arraybackslash}X}
\toprule
Model & OR (95\% CrI) & ICC (95\% CrI) & DIC & WAIC \\
\midrule
CRT-FM-na\"{\i}ve & 0.909 (0.686, 1.204) & -- & 3338.387 & 3338.413 \\
CRT-FM        & 0.971 (0.827, 1.139) & -- & 3328.117 & 3328.100 \\
CRT-MM        & 0.980 (0.726, 1.322) & 0.147 (0.108, 0.188) & 3214.406 & 3205.555 \\
CRT-SMM       & 0.981 (0.726, 1.324) & 0.107 (0.011, 0.174) & 3204.363 & 3200.002 \\
\bottomrule
\end{tabularx}

\begin{tablenotes}
\footnotesize
\item The OR represents the association between children's ITN use and severe symptoms, adjusted for child's age and household wealth index. 
\end{tablenotes}
\end{threeparttable}
\end{table}


\FloatBarrier
\newpage

\FloatBarrier
\newpage
\section*{Supplementary materials}
Supplementary materials are available.
 
\section*{Funding}
No external funding was received for this study.

\section*{Conflict of interest statement}
None declared.

\section*{Data availability}
The R code for implementing the proposed method, simulations, and data examples is available at https://github.com/susanlee505/Spatial-Point-Process-CRT.

\printbibliography
\end{document}